\documentclass[a4paper,12pt]{article}

\usepackage{graphicx,epsf,color,amssymb}

\makeatletter
\renewcommand\section{\@startsection {section}{1}{\z@}%
           {18\p@ \@plus 6\p@ \@minus 3\p@}%
           {9\p@ \@plus 6\p@ \@minus 3\p@}%
           {\normalsize\bfseries\boldmath}}
\renewcommand\subsection{\@startsection{subsection}{2}{\z@}%
           {12\p@ \@plus 6\p@ \@minus 3\p@}%
           {3\p@ \@plus 6\p@ \@minus 3\p@}%
           {\normalfont\normalsize\itshape}}
\renewcommand\subsubsection{\@startsection{subsubsection}{3}{\z@}%
           {12\p@ \@plus 6\p@ \@minus 3\p@}%
           {\p@}%
           {\normalfont\normalsize\itshape}}
\makeatother

\begin{document}

\begin{center}

{\Large\bf Pseudo-Hermitian Position and Momentum Operators, 
     Hermitian Hamiltonian, and Deformed Oscillators}

\vspace*{5mm}

{A.M. Gavrilik$^{a,b}$, I.I. Kachurik$^{a}$}

\vspace*{5mm}

\textit{
${}^a$Bogolyubov Institute for Theoretical Physics of NAS of Ukraine,\\
14-b, Metrolohichna str., Kyiv 03143, Ukraine\\
        ${}^b$omgavr@bitp.kiev.ua }

\begin{abstract}
The recently introduced by us two- and three-parameter ($p,q$)- and
($p,q,\mu$)-deformed extensions of the Heisenberg algebra were
explored under the condition of their direct link with the
respective (nonstandard) deformed quantum oscillator algebras.
 In this paper we explore certain hermitian Hamiltonian build in
 terms of non-hermitian position and momentum operators obeying
 definite $\eta(N)$-pseudo-Hermiticity properties.
    A generalized nonlinear (with the coefficients depending on the
    particle number operator $N$) one-mode Bogoliubov transformation
   is developed as main tool for the corresponding study.
   Its application enables to obtain the spectrum of "almost free"
   (but essentially nonlinear) Hamiltonian.

\end{abstract}

\end{center}

{Keywords}: deformed Heisenberg algebra; position and momentum
  operators; deformed oscillator; $\eta(N)$-pseudo-Hermitian
  operators; Hermitian Hamiltonian; generalized (nonlinear)
  Bogolyubov transformation; energy eigenvalue spectrum

{\it PACS Numbers}: 
       03.65.-w; 03.65.Fd; 02.20.Uw; 05.30.Pr; 11.10.Lm

\newpage

\section{Introduction}


In the last two decades, great deal of attention was devoted to
miscellaneous
 generalizations of the Heisenberg algebra (HA) which use appropriate
 extensions~\cite{Saaved,Brod,Jan,Kempf,Ch-K,GK1,Wess,Plyush}   
  of the relation $ [X, P] = {\rm i}\hbar$.
  Diversity of respective modifications of the famous position-momentum
  uncertainty relation, including those which imply minimal length,
  were also under study, see e.g.~\cite{Jan,Kempf,Garay,Hossen,Q-T,Tkachuk}\,. 

 Recently, in ref.~\cite{GK1} we introduced and studied the
 two-parameter ($p,q$)- and three-parameter ($p,q,\mu$)-deformations of HA.
  In that paper (see also the two subsequent ones \cite{GK2,GK3}),
  the explicit mapping onto certain  
  $q$- or ($p,q$)-deformed oscillator (DO) algebras was obtained.
    If $\mu=0$ and $p=1$, that reduces to the simplest modified HA
    with the $q$-commutator in the l.h.s. of its main relation and
    was studied earlier in \cite{Ch-K}{}.
  Therein, the related "target"  DO algebra (DOA) 
 was obtained as well.
 It is worth noting that there  exists a vast number
 of applications of deformed oscillators or deformed bosons
 (to list a few~\cite{Chang,Liu,AdGa,SIGMA,GKM,GM1,GM2,GM3}).

 In the cases of $q$- and ($p,q$)-deformations of HA, the formulas
 expressing the position and momentum operators $X, P$ in terms of
  the creation and destruction operators $a^+, a^-$, basically differ from
   those (familiar linear ones) for the usual harmonic oscillator,
   as they involve $N$-dependent coefficients,
   with $N$ the excitation number operator.
 Caused by the required realizability of particular deformed HA
 through respective DOA, this fact is of great importance: it suggests
 real distinction from the usual Hermitian conjugation rules
 of the operators $a^-$,\ $a^+$ and others.
  In \cite{GK2}{}, the modified rules of (self)conjugation of the
  involved operators were studied in detail, with different cases listed.
  None of these admits Hermitian rule of conjugation of $a^-$ with
  $a^+$ {\it jointly with} Hermiticity of both position and
  momentum operators. On the other hand, it is possible that usual
  Hermitian conjugation rule is valid for the pair $a^\pm$, but
  $X$ and $P$ are non-hermitian.

  In this paper we study the properties of some hermitian
Hamiltonian built from so-called $\eta(N)$-pseudo-Hermitian position
and momentum operators \cite{GK1}{}.
 As note, the latter inevitably appear if one maps deformed Heisenberg algebra
 onto respective algebra of (nonstandard) deformed oscillator
\cite{Ch-K,GK1}{}.

 Note that non-Hermitian modifications of quantum mechanics which lead nevertheless
 to real spectra of operators, attract great interest, see e.g.
\cite{Geyer,Cannata,Bender,Znojil,Mostafa1,Ahmed,
Mostafa2,Swanson,Bender2,Bender3,Mostafa,Bagchi}{}.
 The approach based on pseudo-Hermitian representation \cite{Mostafa1} in quantum
 mechanics leads to a variety of applications, ranging  \cite{Mostafa} from nonlinear optics
 and nuclear theory, to quantum field theory and even to biophysics.

     In our approach, important role is played by the   
  unusual $\eta(N)$-pseudo-Hermitian conjugation and the related
  $\eta(N)$-pseudo-Hermiticity of $X$ and/or $P$.
   Crucial feature is that the $\eta$-factor which performs
   $\eta(N)$-pseudo-Hermiticity arising within this approach
   due to the mentioned mapping, {\it depends} on the particle number operator $N$.
   That principally differs from the more common pseudo-Hermiticity
   (with $\eta$ depending on momentum) studied e.g.
   in \cite{Mostafa1,Ahmed,Mostafa2,Mostafa,Bagchi}{}.
       Moreover, it was shown in \cite{GK1} that in the case when $a^+$ and $a^-$
   are usual Hermitian conjugates of each other, the both $X$ and
   $P$ should obey such $\eta(N)$-pseudo-Hermiticity.
      Just such situation is in the focus of this paper:
      i.e., we deal with the pair $a^-$,\ $a^+$ which are
    Hermitian conjugates of each other, while $X$ and $P$ are
    $\eta(N)$-pseudo-Hermitian (thus non-hermitian) ones,
    with $N$ the excitation number operator. In terms of these
    operators, we construct an Hermitian Hamiltonian and explore
    some of its properties.

  The paper is organized as follows.
  Section~2 gives a sketch of the deformed version of HA and its
  mapping onto the respective DO (given by a structure function of
  deformation and possessing deformed analog of Fock basis),
  along with inclusion of the  $\eta(N)$-pseudo-hermiticity
  aspects of operators.
    Then, in Sec.~3 an Hermitian Hamiltonian formed
    by $\eta(N)$-pseudo-Hermitian $X$ and $P$ is presented.
  In Sec.~4 we deal with generalized, nonlinearly extended
   Bogoliubov transformation (GNBT) needed to diagonalize the Hamiltonian.
  Section~5 is devoted to the study of unusual (given in the operator form)
   conditions for diagonalization.
     In case of so-called "canonical" GNBT, the spectrum of the diagonalized
     or "almost free" Hamiltonian is obtained in Sec.~7.
    The paper ends with conclusions.

\section{Deformed Heisenberg algebra with $q$- or $p,q$-commutator}

 An approach to modify the Heisenberg algebra by deforming
 commutator was developed in the works~\cite{Wess,Ch-K,
 GK1,GK2,GK3}{}.
 Say, a one-parameter or $q$-deformed analog of HA results if one
 uses the $q$- commutator            
 in the basic or defining relation:  
\begin{equation}                                \label{Kl}
X P-q PX = {\rm i}\hbar \ .
 \end{equation}
  For convenience, in all the treatment below we set $\hbar = 1$.

 Our main requirement, see \cite{Ch-K,GK1,GK2}{}, is that
  the equality (\ref{Kl}) has to be mapped onto some
 DOA whose generating  elements $a^+$, $a^-$ and $N$ -- the
 creation/annihilation operators (not necessarily strict conjugates
 of each other) and the excitation number operator satisfy
\vspace{-1mm}
\begin{equation}                                        \label{N-a}
    [N,a^+]=a^+  \ , \hspace{8mm}  [N,a^-]=-a^- \ ,
\end{equation}
\vspace{-5mm}
\begin{equation}                                         \label{H,G}
H(N)a^- a^+ - G(N)a^+ a^- = 1 \
  \vspace{1mm}
\end{equation}
 where the operator functions $H(N)$ and $G(N)$ admit formal power series
expansion.
  Note that from (\ref{N-a}), for any function
 ${\cal F}(N)$ possessing formal power series expansion, we infer
   \vspace{-1mm}
\begin{equation}                                        \label{F(N)-a}
{\cal F}(N) a^{\pm} = a^{\pm} {\cal F}(N\pm 1), \hspace{7mm}
 [{\cal F}(N),a^\pm a^\mp] =0.
\end{equation}

  Besides (1), we will also consider the two-parameter or
$p,q$-deformed analog~\cite{GK1} of HA whose defining relation is
\vspace{-1mm}
 \begin{equation}                                 \label{qp-H}
q X P -p P X = {\rm i} \hbar \, ,  \hspace{12mm}  p\neq q \, ,    
\end{equation}
where we exclude the trivial $p=q$ case as it reduces to the
standard $[X, P] = {\rm i} \hbar $ \ HA by simple rescaling of $X$
and/or $P$.
 Note that the deformation of HA given in ref.~\cite{Q-T} by the relation
$[X,P]={\rm i}\hbar(1+\alpha^2X^2+\beta^2P^2+\kappa XP+\kappa^*PX)$
  can be at $\alpha=\beta=0$
 related to the $q,p$-algebra (\ref{qp-H})
 through juxtaposing\footnote{(as seen from~\cite{Q-T}{}, nonzero minimal
 uncertainties of $X$ and $P$ do not exist if $\alpha=\beta=0$)}\,:
 $\kappa=\frac{1-q}{{\rm i}\hbar}$ and $\kappa^*=\frac{p-1}{{\rm i}\hbar}$.
 That implies either $p\!=\!q\in {\bf R}$, the trivial excluded case,
 or the other one with $p^*=q=r {\rm e}^{{\rm i}\theta}$.
  In the latter case, one can get rid of the modulus $r$ (again by rescaling
 $X$ and/or $P$). For more details concerning admissible $q$, $p$ in
 (\ref{qp-H}) see~\cite{GK2}{}.

  \subsection{From DHA to DOA}
 We are interested in its mapping onto some DOA, say, given
 by (2)-(3).\
  What is important however, the desired DOA can be as well presented in
  a more useful form by introducing the so-called deformation
 structure function (DSF) $\Phi(N)$,  see e.g. \cite{Melj}{}.
 The latter determines the bilinears
\begin{equation}                                \label{Phi}
a^+ a^- = \Phi(N)\, , \ \ \ \ \   a^- a^+ = \Phi(N+1)\, ,
 \end{equation}
 and hence the commutation relation
 \begin{equation}                                \label{a-,a+}
[a^-, a^+]=\Phi(N+1)-\Phi(N) .
 \end{equation}
 It also determines the action formulas in the
 $\Phi(n)$-deformed analog of Fock space:
   \vspace{-1mm}
\[ N |n\rangle =n |n\rangle, \ \ \
     |n\rangle =\frac{(a^+)^n}{\sqrt{\Phi(n)!}} |0\rangle , \ \ \
  a^-|0\rangle = 0
\]
where $\Phi(n)! \equiv\Phi(n)\Phi(n-1)\ldots \Phi(2) \Phi(1)$ and
also $\Phi(0)!=1$.
 In that space~\cite{Melj,Bona}{},
\begin{equation}                                          \label{a+n}
     \Phi(N) |n\rangle = \Phi(n) |n\rangle\, ,        \hspace{7mm}
 a^+ |n\rangle \!=\! \sqrt{\Phi(n\!+\!1)} |n\!+\!1\rangle ,
 \ \ \ \ a^-|n\rangle \!=\! \sqrt{\Phi(n)} |n\!-\!1\rangle\ .
\end{equation}

 Consider first the $q$-deformed extension of HA.
 The desired DSF can be derived~\cite{GK1}
 if the operator functions $H(N)$, $G(N)$ are known.
  To find these, we set the (nonlinear) relation expressing the
  position/momentum operators through $a^-$,\ $a^+$, namely
\begin{equation}                                         \label{f-g}
X \equiv f(N) a^- + g(N) a^+ , \hspace{11mm}
P \equiv
 {\rm i}\left(
k(N) a^+ - h(N) a^-\right)
\end{equation}
where $f(N)$, $g(N)$, $h(N)$, $k(N)$ are some functions of the
operator $N$.

After simple algebra based on (\ref{f-g}), (\ref{qp-H}) and
(\ref{H,G}) we obtain the expressions
\begin{equation}                                         \label{H(N)}
H(N) = f(N) k(N+1) + q~ h(N) g(N+1) \ ,
\end{equation}
\vspace{-7mm}
\begin{equation}                         \label{G(N)}
G(N) =  g(N) h(N-1) + q~  k(N) f(N-1)\ .
\end{equation}
   For the functions $f, g, h, k$, like in~\cite{Ch-K}
 we have the following relations:
\begin{equation}                                          \label{h/h}
\frac{ h(N\!+\!1)}{ h(N)} = q\, \frac{ f(N\!+\!1)}{ f(N)}\, ,
  \hspace{5mm}
  \frac{ k(N\!-\!1)}{ k(N)} = {q}\, \frac{ g(N\!-\!1)}{ g(N)} \, .
\end{equation}

\subsection{Solutions of the relations (\ref{h/h}) and obtaining the DSF}

 We need the solutions of (\ref{h/h}) which then, using (\ref{H(N)}) and (\ref{G(N)}),
 yield the corresponding operator functions $G(N)$ and $H(N)$.
  Such solutions were found in \cite{GK1,GK3}{}, and here we will deal with
  the following two of them.

\vspace{1mm}
  \underline{Solution with single parameter $q$}.
  This solution implies
\begin{equation}                         \label{f,k(N)}
f(N)\!=\! k(N)\!=\!\frac{1}{\sqrt 2}\, {q}^{N},
      \hspace{7mm}
 h(N)\!=\! g(N)\!=\! \frac{1}{\sqrt 2} \, {q}^{2N} \ ,
\vspace{-2mm}
\end{equation}
 from which
 \begin{equation}                      \label{H(N)q}
 H(N) = \frac12\, {q}^{2N+1}\Bigl(1
 + q^{2N+2} \Bigr)\,  ,           \hspace{10mm}
G(N) = \frac12\, {q}^{2N}\Bigl(1 + {q}^{2N-2} \Bigr)
 = \, q^3 H(N-2) .
\end{equation}
Besides, putting (\ref{f,k(N)}) into (\ref{f-g}), 
for the operators $X$ and $P$ we obtain:
 \begin{equation}                                             \label{X,P_q'N}
X\!=\!\frac{1}{\sqrt 2} \, \Bigl(q^{2N} a^+ +  q^N a^- \Bigr) ,
\hspace{3mm}
 P\!=\!\frac{\rm i}{\sqrt 2} \, \Bigl(q^N a^+\!-\!q^{2N} a^-\Bigr).
 \end{equation}
Using $H(N)$ and $G(N)$ from (\ref{H(N)q}), the explicit expression
for the related DSF can be obtained, see \cite{Ch-K,GK1}{}.
 We will give that DSF at the end of this subsection.

\vspace{1mm}
 \underline{Solution with two parameters $q, p$}.   
 For the $q,p$-deformed case the operators $X$ and $P$ are expressed as~\cite{GK3}
   \begin{equation}
 \vspace{-1mm}                                    \label{X,P_Q'N}
X = \frac{1}{\sqrt 2} \, \Bigl[Q^{2N}\!a^+
 +  Q^N\!a^- \Bigr]  ,  \hspace{7mm}
P = \frac{\rm i}{\sqrt 2} \,
 \Bigl[
Q^N\!a^+ -
  Q^{2N}\!a^-\Bigr] ,     \hspace{8mm}          Q\equiv q/p .
\end{equation}
 The inverse relations readily follow, namely
\begin{equation}
 \vspace{-1mm}                                    \label{a_vs_X}
a^-\!=\! d_{N,Q}\bigl(Q^{-N}X\!+ {\rm i} P\bigr)   , \hspace{5mm}
a^+\!=\! d_{N,Q}\bigl(X\!- {\rm i} Q^{-N} P\bigr), \hspace{6mm}
d_{N,Q}\equiv\sqrt2 (1+Q^{2N})^{-1}.
\end{equation}
 Obviously, at $p=1$ the latter relations give the (inverse) formulas
 for the $q$-deformed case. Further restriction $q=1$
 implies $d_{N,1}=\frac{1}{\sqrt2}$ and recovers usual linear
 relations (see e.g. \cite{Davydov}) between  $a^+, a^-$ and $X, P$.

 In the two-parameter case of $q,p$-deformed HA
 the operator functions $H(N)$ and $G(N)$ were also found~\cite{GK3}
 that allowed to obtain the desired DSF for (\ref{a-,a+}):
\[  
    \hspace{14mm}    
 \Phi_{p,q}(n) \!=\! \frac{2 p^{-1} Q^{-n}}{(1+Q^{2n-2}) (1+Q^{2n})}
 \left(1\!+\!\frac{Q^{n}\!-\!
  Q^{-n+1}}{Q-1}\right) =
  \] 
\begin{equation}                                           \label{Phi'(n)}
 \ \ \ =\! \frac{2 q^{-n} p^{5n-3}}
                {(q^{2n-2}+p^{2n-2})(q^{2n}+p^{2n})}
 \left(1\!+\!\frac{[2n\!-\!1]_{q,p}}{(qp)^{n-1}}\right)\, .
\end{equation}
Here, $[m]_{q,p}\equiv \frac{q^m-p^m}{q-p}$ is the $q,p$-number
corresponding to a number $m$, and the relation $\Phi(N)|n\rangle =
\Phi(n)|n\rangle$ for the DSF, see eq.~(\ref{a+n}), has been used.

    Formula (\ref{Phi'(n)}) gives the DSF of nonstandard
two-parameter deformed quantum oscillator.
 "Nonstandard" means nonsymmetric under $q\leftrightarrow p$ because
 of the factor $q^{-n}p^{5n-3}$ in the numerator.
 Due to that it obviously differs from the well-known $q,p$-oscillator \cite{Chakra}
 whose structure function $\varphi_{q.p}(n)=[n]_{q,p}$ is
  ($q\leftrightarrow p$)-symmetric.

  At last, the one-parameter or $q$-deformed DSF follows from the $p,q$-deformed
  one in (\ref{Phi'(n)}) as special case if we set $p=1$.
     That is,
\begin{equation}                                           \label{Phi(n)}
{\Phi_q(n)} =  \frac{2 q^{-n}}{(1\!+\!q^{2n-2})
(1\!+\!q^{2n})}\Bigl(1\!+\!\frac{q^{n}\!-\!q^{-n+1}}{q\!-\!1}\Bigr)
 = \frac{2\,q^{-n} \,[n]_q \,(1+q^{-n+1}) }{(1+q^{2n-2}) (1+q^{2n})}
\,
 \end{equation}
  where $[n]_q\equiv (1-q^n)/(1-q)=\phi_q(n)$ (the latter coincides with
  the DSF of Arik-Coon deformed oscillator~\cite{AC}).
 The obtained DSFs (16),(\ref{Phi(n)}) imply that now we have,
 besides the relation (\ref{H,G}), also the commutation relation in
 the alternative form (\ref{a-,a+}).

 Note that from the two-parameter family with DSF in~(\ref{Phi'(n)})
 one can infer, by imposing different functional relations $p\!=\!f(q)$
 similarly to~\cite{Plethora}{}, a "plethora" of one-parameter DOs.

\section{Mutually conjugate $a^+$, $a^-$, and
 $\eta(N)$-pseudo-Hermitian position and momentum operators}

 We require that $a^+$ and $a^-$ obey the customary conjugation
property:
 \begin{equation}                  \label{a^mp}
 (a^\pm)^\dagger =  a^\mp .
 \end{equation}
 Then as shown in \cite{GK2} it follows that
both $X^\dagger\neq X$ and $P^\dagger\neq P$, and one of the
possibilities is to consider these operators as
$\eta(N)$-pseudo-Hermitian ones of the form
 \begin{equation}                           \label{X_eta}
X^\dagger = \eta^{-1}_{\scriptstyle{X}}(N) \, X \,
\eta^{}_{\scriptstyle{X}}(N) \, ,  \hspace{10mm}
  P^\dagger = \eta^{-1}_{\scriptstyle{P}}(N)   \, P \, \eta^{}_{\scriptstyle{P}}(N).
\end{equation}
 In ref.~\cite{GK2}{}, $\eta^{}_{\scriptstyle{X}}(N)$ and
 $\eta^{-1}_{\scriptstyle{P}}(N)$ were found explicitly, by exploiting
 certain recurrence relations
\[
\eta^{}_{\scriptstyle{X}}(N+1)=\eta^{}_{\scriptstyle{X}}(N)\,
Q^{N+2} \, , \hspace{10mm}
\eta^{}_{\scriptstyle{P}}(N+1)=\eta^{}_{\scriptstyle{P}}(N)\,
Q^{-N+1}\, .
\]
Solving of these yields
\[
\eta^{}_{\scriptstyle{X}}(N) = \
 Q^{\frac12 N(N+3)}\, \eta^{}_{\scriptstyle{X}}(0)\, , \hspace{10mm}
 \eta^{}_{\scriptstyle{P}}(N) =
\, Q^{\frac12 N(-N+3)}\, \eta^{}_{\scriptstyle{P}}(0) ,
\]
and the convenient choice is to set $\eta^{}_{\scriptstyle{X}}(0)=
\eta^{}_{\scriptstyle{P}}(0)=1$.
 As result, we have
\begin{equation}                                \label{X,P_Q'N}
X^\dagger = Q^{-\frac12 N(N+3)}\, X \, Q^{\frac12 N(N+3)}\, ,
\hspace{10mm}
  P^\dagger = Q^{\frac12 N(N-3)}\, P \, Q^{-\frac12 N(N-3)}\,.
\end{equation}
Although $X$ and $P$ are $\eta(N)$-pseudo-Hermitian (i.e.
non-Hermitian of special form), in terms of these non-Hermitian
operators we can nevertheless construct Hermitian Hamiltonian(s),
see below.

 \underline{Remark 1}. Let us note that, as discussed in ref.~\cite{GK2}{},
 besides the considered case (\ref{a^mp}) of $a^{\pm}$ being mutual conjugates,
there also exist the cases in which the operators $a^+$,\ $a^-$ are
$\eta_a(N)$-pseudo-Hermitian conjugates of one another.
 For those cases, only one (or none) of the position/momentum operators
can be Hermitian.

     \section{Hermitian Hamiltonian from non-Hermitian $X, P$}

 Simplest choice is to take the Hamiltonian in the familiar form
 $H =\frac12 \bigl(a^- a^+ + a^+ a^- \bigr)$ where we have set $\hbar\omega =1$.
   In view of (\ref{a^mp}), this form of Hamiltonian guarantees its Hermiticity.
    With account of the equality
\[
\frac{p}{2} Q^{2N+1} \bigl(1+Q^{2N+2}\bigr) a^-a^+ - \frac{p}{2}
Q^{2N} \bigl(1+Q^{2N-2}\bigr) a^+ a^- = 1
\]
 which follows from (\ref{H,G}) and the expressions (\ref{H(N)q})
 for $H(N)$, $G(N)$, the Hamiltonian $H$ can be presented in the form
\begin{equation}                       \label{H_by_a2}
 H = \frac{1}{p} \ \frac{Q^{-2N-1}}{1+Q^{2N+2}} +\frac12
 \biggl(1+ Q^{-1}\frac{1+Q^{2N-2}}{1+Q^{2N+2}}
\biggr) a^+ a^- \, .
\end{equation}
 Besides that it is Hermitian, we can easily write down its energy
 spectrum $E_n=\frac12\left(\Phi(n+1)+ \Phi(n)\right)$,
 by the account of DSF $a^+ a^-=\Phi(N)$ from eq.~(\ref{Phi})
 applied to the Fock basis state $\vert n \rangle$, and taking the
 expression (\ref{Phi(n)}) for $\Phi_{q}(n)$.
When $p=Q=1$, the usual harmonic oscillator
 with ${\cal H} = {\cal H}_0 = \frac12 + a^+ a^-=N+\frac12$ is recovered.

 However, we are interested in the nontrivial Hamiltonian containing
 the $\eta(n)$-pseudo-Hermitian position and momentum operators considered above.
  It is clear that the familiar form $H=\frac12 (X^2+P^2)$ is not
  admissible in our situation, being neither Hermitian nor pseudo-Hermitian
   in the deformed case (i.e. for $Q\neq 1$).
  That is why we suggest a natural and simple modification of
  $H$ which involvse the $\eta_X$-pseudo-Hermitian operator $X$ and
  $\eta_P$-pseudo-Hermitian operator $P$, namely
\begin{equation}                       \label{H_XP_eta}
 {\cal H} = \frac12 \Bigl( (\eta_X)^{-\frac12} X^2 (\eta_X)^{\frac12}
+  (\eta_P)^{-\frac12} P^2  (\eta_P)^{\frac12} \Bigr) ,
\end{equation}
whose Hermiticity follows from (\ref{X_eta}).

With the explicit $\eta_X(N)$ and $\eta_P(N)$ related
 (see~\cite{GK2} and Subsec.~2.3) with the case $(a^\pm)^\dagger=a^\mp$, we have
\begin{equation}                       \label{H_XP_herm}
 {\cal H}_Q =
 \frac12 \Bigl( Q^{-\frac14 N(N+3)} X^2 \, Q^{\frac14 N(N+3)} +
 Q^{\frac14 N(N-3)} P^2 \, Q^{-\frac14 N(N-3)} \Bigr) \ .
 \end{equation}
 Using (\ref{X,P_Q'N}), we easily verify Hermiticity of ${\cal H}_Q$.
    Note that if $Q\to 1$, we recover $H =\frac12(X^2 + P^2)$.
 Below, the Hamiltonian (\ref{H_XP_herm}) with $Q\to q$ and the
 related DSF (\ref{Phi(n)}) will be main subject of our study.

\begin {center}
\underline{\it Hermitian Hamiltonian ${\cal H}$ in (\ref{H_XP_herm})
as nonlinear analog of Swanson model}
\end {center}

\noindent
 From (\ref{X,P_q'N}) we have
 \[                                  \hspace{-12mm}
 X^2= \frac12 (q^{4N-2} a^+ a^+ + q^{3N-1} a^+ a^- + q^{3N+2} a^- a^+
+ q^{2N+1} a^- a^-)\ ,
 \]
\begin{equation}   \hspace{7mm}                          \label{X'2,P'2}
P^2= -\frac12(q^{2N-1} a^+ a^+ - q^{3N-2} a^+ a^- - q^{3N+1} a^- a^+
+ q^{4N+2} a^- a^-)\ .
\end{equation}
 Plugging these in ${\cal H}_q$ in eq.~(\ref{H_XP_herm})
  we arrive at the (nonlinear, non-diagonal) {\it Hermitian} Hamiltonian
  expressed through the annihilation/creation operators, namely
\begin{equation}                           \label{H_sw}
{\cal H}_q = A_q(N) a^+ a^+ + B_q(N) a^- a^- + C_q(N) a^+ a^- +
D_q(N) a^- a^+
\end{equation}
where
\[
  {A}_q(N) = \frac14 q^{3N-3}\left(q^{1/2}- q^{-1/2}\right)\ ,  \hspace{12mm}
  {B}_q(N) = \frac14 q^{3N+3}\left(q^{1/2}- q^{-1/2}\right)\ ,
\]
\[
\ \ \
 {C}_q(N) = \frac14 q^{3N-\frac32}\left(q^{1/2} + q^{-1/2}\right)\ , \hspace{12mm}
  {D}_q(N) = \frac14 q^{3N+\frac32}\left(q^{1/2} + q^{-1/2}\right)\ .
\]
 The obtained Hamiltonian is reminiscent of the Swanson's
 model~\cite{Swanson} due to the presence of $a^+ a^+$ and $a^- a^-$ terms.
 However, Swanson's Hamiltonian is non-Hermitian because of the
 differing {\it coupling constants} $\alpha$,\ $\beta$ in front of $a^+ a^+$ and $a^- a^-$
 ($a^+$ and $a^-$ were taken in~\cite{Swanson} as usual boson operators).
  On the other hand, in our Hamiltonian (which is Hermitian) we have,
  instead of numerical constants,the
  {\it operator functions} as "coefficients" in front of $a^+ a^+$ and $a^- a^-$.
  Moreover, we deal with $a^+$ and $a^-$ describing {\it deformed} bosons.

 \underline{Remark 2}.
 Notice that $A_q(N)\neq B_q(N)$ in (\ref{H_sw}).
 On the contrary, if we had $A_q(N)=B_q(N)$, the Hamiltonian could not be
 Hermitian because of relations (\ref{a^mp}) and (\ref{F(N)-a}).
 Fortunately, the explicit form of $A_q(N)$, $B_q(N)$ shows they are
 unequal, and related as
  \[
 {B}_q(N) = q^{6} {A}_q(N)\, , \hspace{12mm}  {D}_q(N) = q^{3} \, {C}_q(N)\, .
\]
 Just this relation of proportionality of $A_q(N)$ and $B_q(N)$ provides
 the Hermiticity of ${\cal H}_q$ for real $q$, while ${C}_q(n)$ and ${D}_q(n)$
 may  
 be any real functions. In the case of phase-like $q$
 i.e.  $q={\rm e}^{{\rm i}\theta}$, detailed analysis shows that for such $q$
 (and for general complex $q$) the Hamiltonian (\ref{H_sw})
 cannot be Hermitian. So $q$ must be real, $q>0$.

 Let us also note that in the no-deformation limit $q\to 1$, due to
 $A_q(N)|_{q\to 1} \to 0$ and $B_q(N)|_{q\to 1} \to 0$ (see (\ref{H_sw})\,),
 the terms containing $a^+a^+$ and $a^-a^-$ disappear from the Hamiltonian.
 That is, in our case the terms with $a^+a^+$ and $a^-a^-$ exist
 just {\it due to deformation}. In a sense, nontrivial deformation
 ($q\neq 1$) in our case corresponds to non-vanishing $\alpha$,\, $\beta$
 in Swanson's Hamiltonian.

\section{Example of $\eta(N)$-pseudo-Hermitian Hamiltonian}

It is natural that in terms of $\eta_X(N)$-pseudo-Hermitian position
 and $\eta_P(N)$-pseudo-Hermitian momentum operators, see (\ref{X_eta}),
 one can easily construct an $\eta(N)$-pseudo-Hermitian Hamiltonian.
  Here we will present a rather simple, pseudo-Hermitian Hamiltonian which is
  very similar to the above Hermitian Hamiltonian.

  Under the same conditions as above, i.e. for
 $(a^\pm)^\dagger=a^\mp $ along with $X$, $P$ from (\ref{X,P_Q'N})
 we obtain the Hamiltonian
\begin{equation}                               \label{H_psh}
\tilde{H}_{{}_{\rm PsH}} = \frac12 \left(Q^{\frac14 N(3-N)} X^2\,
 Q^{\frac14 N(N-3)} + Q^{\frac14 N(N+3)} P^2 \, Q^{-\frac14 N(N+3)}\right) .
\end{equation}
 One can easily verify that this is $\eta_{H}(N)$-pseudo-Hermitian
 with $\eta_{H}(N)=Q^{3N}$.

 \underline{Remark 3}.
  It is worth noting that the above $\eta_H(N)$-pseudo-hermitian
 Hamiltonian is very similar to the Hermitian Hamiltonian given in eq.~(\ref{H_XP_herm}).
 Moreover, from the $\eta_H(N)$-pseudo-hermitian Hamiltonian~(\ref{H_psh}) one can
 formally obtain the Hermitian one~(\ref{H_XP_herm}) by the composition of two
 exchanges: $X \leftrightarrow P$ and then $q\rightarrow q^{-1}$.

\vspace{1mm}
 \underline{Remark 4}.
 The Hamiltonian eq.~(\ref{H_psh}) can be presented in an almost "standard" form.
 Indeed, denoting $ \tilde{X}= Q^{-\frac{N^2}{4}} X\,
Q^{\frac{N^2}{4}}$ and $\tilde{P}= Q^{\frac{N^2}{4}} P\,
Q^{-\frac{N^2}{4}}$, we arrive at
\begin{equation}
 {H}_{{}_{\rm PsH}} = \frac12 Q^{\frac{3N}{4}}\bigl( \tilde{X}^2
 +  \tilde{P}^2 \bigr) Q^{-\frac{3N}{4}}\,  \ \ \ \ \  {\rm or} \ \ \
 \ \
\tilde{{\cal H}} = \frac12 \bigl(\tilde{X}^2
 +  \tilde{P}^2 \bigr)
\end{equation}
where $\tilde{{\cal H}}\equiv  Q^{-\frac{3N}{4}} H_{{}_{\rm PsH}}\,
Q^{\frac{3N}{4}} $.
  Note also that while $X$ is $\eta_X(N)$-pseudo-Hermitian of the form
\[
X^\dagger = \eta_X^{-1}(N)X\, \eta_X(N) = Q^{-\frac12 N(N+3)} X\,
Q^{\frac12 N(N+3)},
\]
its tilted counterpart satisfies:
\[   \tilde{X}^\dagger = Q^{-\frac{3N}{2}} \tilde{X}\,
Q^{\frac{3N}{2}} \, ,
\]
what means $\tilde{X}$ is $\eta_{\tilde{X}}(N)$-pseudo-Hermitian
with $\eta_{\tilde{X}}(N)=Q^{\frac{3N}{2}}$.

Likewise, while $P$ is $\eta_P(N)$-pseudo-Hermitian of the form
\[
P^\dagger = \eta_P^{-1}(N) P\, \eta_P(N) = Q^{\frac12 N(N-3)} P\,
Q^{-\frac12 N(N-3)},
\]
its tilted counterpart satisfies:
\[   \tilde{P}^\dagger = Q^{-\frac{3N}{2}} \tilde{P}\,
Q^{\frac{3N}{2}} \, ,
\]
what means $\tilde{P}$ is $\eta_{\tilde{P}}(N)$-pseudo-Hermitian
with $\eta_{\tilde{P}}(N)=Q^{\frac{3N}{2}}$.

\vspace{1mm}
\begin {center}
\underline{\it Hamiltonian ${H}_{\rm PsH}$ in (\ref{H_psh}) as
nonlinear analog of Swanson model}
\end {center}

  Like in Hermitian case, using eq.~(\ref{X'2,P'2}) the $\eta(N)$-pseudo-Hermitian
  Hamiltonian (\ref{H_psh}) can also be presented in terms of
  annihilation/creation operators.
  Indeed,    
\begin{equation}                                  \label{{H_sw'}}
 {H}_{{}_{\rm PsH}} =
  \tilde{A}_N a^+ a^+ + \tilde{B}_N a^- a^- + \tilde{C}_N a^+ a^- + \tilde{D}_N a^-
a^+ \ ,
\end{equation}
\[
 \  \tilde{A}_N = \frac14 q^{3N} \left(q^{1/2} - q^{-1/2}\right)  \ ,  \hspace{16mm}
  \tilde{B}_N = \frac14 q^{3N} \left(q^{1/2} - q^{-1/2}\right) \ ,
\]
\[
\ \ \
 \tilde{C}_N = \frac14 q^{3N-\frac32} \left(q^{1/2} + q^{-1/2}\right) \ , \hspace{11mm}
  \tilde{D}_N = \frac14 q^{3N+\frac12} \left(q^{3/2} + q^{-3/2}\right)
\]
(compare with eq.~(\ref{H_sw})\,). Notice that now $\tilde{A}_N =
\tilde{B}_N $,\ $\tilde{C}_N = q^{-2} \tilde{D}_N$.   
 Detailed study of  $\eta(N)$-pseudo-Hermitian Hamiltonian
 (the spectrum etc.) will be done elsewhere.

 \section{General nonlinear Bogoliubov transformation}

Basically we intend to find spectra of eigenvalues of the both
Hamiltonians (27) and (30). However, since the treatment of
non-Hermitian Hamiltonian is much more involved than Hermitian case,
in the rest of this paper we restrict ourselves to the study of the
Hermitian Hamiltonian, see Section 7 below.

In order to perform diagonalization of the Hermitian Hamiltonian
(27), we have first to study general nonlinear Bogoliubov
transformations (GNBT) between any two deformations of the quantum
oscillator (note that some generalizations of Bogolyubov
transformation involving deformed bosons were studied earlier
in~\cite{Gango,Katriel,Zhed,Naderi}\,).
  To this end, we introduce the new pair of creation/dectruction
operators defined as
\begin{equation}
c = g_1{(N)} a^+ +  g_2{(N)} a^- \ ,
\end{equation}
                  \vspace{-7mm}
\begin{equation}
d = g_3{(N)} a^+ +  g_4{(N)} a^- \ .
\end{equation}
Nonlinearity of these GNBT stems from the non-constant nature of the
(operator) coefficient functions $g_i(N)$ of the Hermitian operator
$N=\Phi^{-1}(a^+ a^-)$, see eq.~(\ref{Phi}).

We require $c$ and $d$ to be mutual Hermitian conjugates of one
another so that
\[
c^\dagger = d  \hspace{10mm}
 \Longrightarrow  \hspace{9mm}  g_3{(N)} = g_2{(N-1)}
, \ \
 g_4{(N)} = g_1{(N+1)}\
 \]
 and thus
\begin{equation}                        \label{c_a^pm}
\hspace{-14mm} \ \ \ c = g_1{(N)}\, a^+ +  g_2{(N)}\, a^- \ ,
\end{equation}
\begin{equation}                          \label{c+_a^pm}
c^\dagger = g_2{(N\!-\!1)}\, a^+ +  g_1{(N\!+\!1)}\, a^- \ .
\end{equation}
In the matrix form that reads
\[
\left(\matrix{ c    \cr
                     c^\dagger\cr
       }\right) =
\left(\matrix{\hspace{-6mm} g_2(N)  & \hspace{-3mm} g_1(N) \cr
                   \, g_1(N\!+\!1)  & \ \  g_2(N\!+\!1) \cr
       }   \right)          \!
       \left(\matrix{ a^- \cr
                            a^+ \cr
      }\right)
\]
where the elements of matrix are (mutually commuting) operator
functions of $N$.

\noindent
 With the notation $c^\dagger\equiv c^+$, \ $c\equiv c^-$,
the inverse of (\ref{c_a^pm}), (\ref{c+_a^pm}) is           
\begin{equation}              \label{a-c^pm}
 a^- = {\cal K}_N \left( g_1(N)\, c^+ - g_2(N\!-\!1)\, c^- \right) ,
\end{equation}
\begin{equation}            \label{a+_c^pm}
a^+ = {\cal K}_N \left( g_1(N\!+\!1)\, c^- - g_2(N)\, c^+ \right) \
\end{equation}
where
\[
  {\cal K}_N\equiv  \Bigl( g_1(N) g_1(N\!+\!1) - g_2(N) g_2(N\!-\!1)\Bigr)^{-1}.
\]
Now assume that, after applying GNBT to the couple ($a^+$, $a^-$) we
are led to another deformed operators $c^+$ and $c^-$ obeying the
relations\footnote{The resulting commutator is unchanged if we
replace $\chi(N)\to \chi(N)+\psi$ where $\psi=\psi({\rm def.par.})$
is some function of deformation parameter(s) only.
  For simplicity, we will drop it.}
\begin{equation}          \label{chi}
[c^-,c^+] = \chi(N+1) - \chi(N)\ , \hspace{14mm} c^+ c^- = \chi(N) .
\end{equation}
From (\ref{chi}) and (\ref{c_a^pm})-(\ref{c+_a^pm}), by simple
algebra we
 infer
\[  \hspace{-12mm}
c^- c^+ - c^+ c^- =
 \left[ g_1(N)g_2(N\!-\!2)-g_2(N\!-\!1)g_1(N\!-\!1)\right]a^+a^+ +
\]
\[   \hspace{-19mm}
 +\left[ g_1^2(N)-g_2^2(N\!-\!1)\right]a^+a^- +
\]
\[   \hspace{-19mm}
 +\left[ g_2^2(N)-g_1^2(N\!+\!1)\right]a^-a^+ +
\]
\[ \hspace{12mm}
+\left[g_2(N)g_1(N\!+\!2)-g_1(N\!+\!1)g_2(N\!+\!1)\right]a^-a^-\ .
 \]
 Clearly, validity of (\ref{chi}) imposes the following conditions
\[ g_1(N)g_2(N\!-\!2)-g_2(N\!-\!1)g_1(N\!-\!1)= 0 ,
\]
 \[
 g_2(N)g_1(N\!+\!2)-g_1(N\!+\!1)g_2(N\!+\!1)=0 ,
  \]
  \[
g_1^2(N)-g_2^2(N\!-\!1)= - \chi(N)/\Phi(N) , \hspace{12mm}
 g_2^2(N)-g_1^2(N\!+\!1) = \chi(N+1)/\Phi(N+1) \, .
 \]
 The first two relations are equivalent (with shift $N\to N+2$).
Likewise equivalent (with shift $N\to N+1$) are the last two.
 Hence we have two independent relations:
\begin{equation}                    \label{frac}
\frac{g_1(N\!+\!1)}{g_2(N)} = \frac{g_1(N)}{g_2(N-1)} =
\epsilon({\rm d.p.}) , \hspace{12mm}   g_1^2(N)-g_2^2(N\!-\!1)= -
 \chi(N)/\Phi(N) \,
\end{equation}
where $\epsilon({\rm d.p.})\equiv\epsilon({\rm def.par.})$ reflects
the fact that the ratio does not depend on $N$, but may depend on
deformation parameter(s) involved in the structure functions
$\chi(N)$, $\Phi(N)$.
 From the ratio in (\ref{frac}) we have  $g_2(N\!-\!1) = \epsilon^{-1} g_1(N)$.
 Then, the explicit formulas for $g_1(N)$ and $g_2(N)$ do follow:
\begin{equation}                         \label{gen_g1g2}
g_1(N)=\epsilon \,\sqrt{\frac{\chi(N)}{(1\!-\!\epsilon^2)\,\Phi(N)}}
, \hspace{9mm}
g_2(N)=\sqrt{\frac{\chi(N+1)}{(1\!-\!\epsilon^2)\,\Phi(N+1)}} .
\end{equation}
Let us note that if $\epsilon=\pm 1$ the formula (38) gives $\chi
=0$ which  in view of (37) implies $c^+c^- = c^-c^+ =0$. From this
and (33) we conclude that then $g_1=g_2=0$, and the whole concept of
GNBT loses its sense. Hence, from now on $\epsilon\neq\pm 1$.

The obtained operator functions $g_1(N)$ and $g_2(N)$ in
(\ref{gen_g1g2}) provide most general (single-mode) nonlinear
Bogoliubov transformation (\ref{c_a^pm})-(\ref{c+_a^pm}) from the
deformed-boson operators $a^+$,\ $a^-$ (or $\Phi$-oscillator) to
another deformed boson operators $c^+$,\ $c^-$ ($\chi$-oscillator)
such that, denoting $\zeta(N) \equiv \sqrt{\Phi(N)/\chi(N)}$, we
have

\begin{equation}              \label{c-}
c^- = \frac{1}{\sqrt{1\!-\!\epsilon^2}}
 \Bigl( \, \epsilon\,\zeta^{-1}(N) \,           
a^+ + \zeta^{-1}(N\!+\!1)\,a^- \Bigr) ,         
 \end{equation}
 \vspace{-4mm}
\begin{equation}
 c^+ =                         \label{c+}
\frac{1}{\sqrt{1\!-\!\epsilon^2}}
 \Bigl(\,\zeta^{-1}(N)\, a^+ +
 \epsilon\,\zeta^{-1}(N\!+\!1)\,a^-\Bigr) .
\end{equation}
 In the matrix form this looks as
\[
\left(\matrix{ c^-    \cr
                     c^+\cr
       }\right) =
  || \hat{A}|| \left(\matrix{ a^- \cr
                            a^+ \cr
        }\right) \equiv
\left(\matrix{
     \hspace{1mm} \frac{\zeta^{-1}(N+1)}{\sqrt{1\!-\!\epsilon^2}}
       & \hspace{-3mm} \ \ \
\frac{\epsilon\,\zeta^{-1}(N)}{\sqrt{1\!-\!\epsilon^2}}    \cr
     \frac{\epsilon\,\zeta^{-1}(N+1)}{\sqrt{1\!-\!\epsilon^2}}  &
                   \ \ \frac{\zeta^{-1}(N)}{\sqrt{1\!-\!\epsilon^2}}  \cr
       }   \right)          \!
       \left(\matrix{ a^- \cr
                            a^+ \cr
       }\right) ,  \hspace{6mm}
       \det \hat{A} = \frac{1}{\zeta(N)\,\zeta(N\!+\!1)} .
\]
The operators $c^+$ and $c^-$ are mutual conjugates,  
and the requirement $\epsilon\neq\pm 1$ implies that $-1 <\epsilon <
1$. In case of $\epsilon=0$, the matrix takes diagonal form, with
function of $N$ on the diagonal.  
 This case looks similar to the familiar transformation~\cite{Bona}
 from 
 nondeformed (or bosonic) creation/destruction operators to
 their deformed counterpart.

The inverse transformation reads
\begin{equation}                               \label{inverse}
 a^- = 
  \frac{\zeta(N+1)}{\sqrt{1\!-\!\epsilon^2}}
    \left( c^- - \epsilon\, c^+  \right) , \hspace{6mm}
 a^+ =  
   \frac{\zeta(N)}{\sqrt{1\!-\!\epsilon^2}}
 \left(c^+ -\epsilon\, c^- \right) ,
\end{equation}
or
\begin{equation}                               \label{inverse}
\left(\matrix{ a^-    \cr
                     a^+\cr
      }\right) =
\left(\matrix{
     \hspace{1mm} \frac{\zeta(N+1)}{\sqrt{1\!-\!\epsilon^2}}   &
      \hspace{-3mm} \ \ \ \frac{-\,\epsilon\,\zeta(N+1)}{\sqrt{1\!-\!\epsilon^2}}\cr
                   \frac{-\,\epsilon\,\zeta(N)}{\sqrt{1\!-\!\epsilon^2}}  &
                   \ \ \frac{\zeta(N)}{\sqrt{1\!-\!\epsilon^2}} \cr
       }   \right)          \!
       \left(\matrix{ c^- \cr
                            c^+ \cr
       }\right) ,  \hspace{9mm}
       \det \hat{A}^{-1} = \zeta(N)\,\zeta(N+1) .
\end{equation}

 \underline{Remark 5}.
 The nonlinear Bogolyubov transformation that preserves
commutation relation up to some constant multiplier $\kappa_q\equiv
\kappa(q)$ so that
\begin{equation}         \label{kappa}
   \chi(N) = \kappa_q\, \Phi(N)\, ,  \hspace{14mm}
 \lim_{q\to1}\kappa_q =1\, ,
\end{equation}
will be called {\it canonical} GNBT.
 In this case we have $\zeta(N)=\kappa_q^{-1/2}$, i.e. zeta is a the constant.
  In particular, $\kappa_q$ may be put equal to 1 that implies $\zeta(N)=1$.
  Recall that usual canonical Bogolyubov transformation transforms
  bosons into bosons.
  In the canonical case, the GNBT and their inverse read:
  \begin{equation}                             \label{can}
      c^- = 
  \frac{\kappa_q^{1/2}}{\sqrt{1\!-\!\epsilon^2}}
    \left( \epsilon\,a^+ + a^-   \right) , \hspace{6mm}
 c^+ =  
   \frac{\kappa_q^{1/2}}{\sqrt{1\!-\!\epsilon^2}}
 \left(a^+ +\epsilon\,a^- \right) ,
   \end{equation}
\vspace{-5mm}
   \begin{equation}                       \label{can-inverse}
 a^- = 
  \frac{\kappa_q^{-1/2}}{\sqrt{1\!-\!\epsilon^2}}
    \left( c^- - \epsilon\,c^+\right) , \hspace{6mm}
 a^+ =  
   \frac{\kappa_q^{-1/2}}{\sqrt{1\!-\!\epsilon^2}}
 \left(c^+ -\epsilon\,c^- \right) .
 \end{equation}

\vspace{0mm}
 \section{ Hamiltonian for general case of quasi-particles}

Now let us go back to the Hermitian Hamiltonian (\ref{H_sw}).
 With the account of (\ref{inverse}), the Hamiltonian is expressed through the
 new deformed operators $c^+$ and $c^-$, i.e.
\[
{\cal H}_{q,\epsilon} =
 \frac{1}{1\!-\!\epsilon^2} R_q(N)\, c^+\,\zeta(N)\,c^+
 + \frac{\epsilon}{1\!-\!\epsilon^2} S_q(N)\, c^+\zeta(N+1)\, c^+
\]
\[
 + \frac{\epsilon}{1\!-\!\epsilon^2} T_q(N)\, c^- \, \zeta(N)\, c^-
 +
 \frac{1}{1\!-\!\epsilon^2} U_q(N)\,c^-\zeta(N+1)\,c^-
\]
\[
  \hspace{0mm}   - \frac{\epsilon}{1\!-\!\epsilon^2} R_q(N)\, c^+\, \zeta(N)\, c^-
    - \frac{1}{1\!-\!\epsilon^2} S_q(N)\, c^+\, \zeta(N+1)\, c^-
 \]
\begin{equation}                                      \label{H'c+c-}
 - \frac{1}{1\!-\!\epsilon^2} T_q(N) \, c^- \, \zeta(N)\, c^+
 - \frac{\epsilon}{1\!-\!\epsilon^2} U_q(N)\, c^-\, \zeta(N+1)\, c^+
\end{equation}
where
\[                                      
R_q(N)= A_q(N) \, \zeta(N) - \epsilon D_q(N) \zeta(N+1)\ ,
 \hspace{5mm}
 S_q(N)=  \epsilon B_q(N) \zeta(N+1)
 - C_q(N) \zeta(N) \ ,
\]                                   
 \[                                 
 T_q(N)= \epsilon A_q(N) \zeta(N) -
  D_q(N) \zeta(N+1)\ , \hspace{5mm}
  U_q(N)= B_q(N) \zeta(N+1)
 - \epsilon\,C_q(N) \zeta(N) \ .
 \]                                  

In the limit $q\to 1$ we have $R_q(N)\,\to
-\frac{\epsilon}{2}\tilde{\zeta}(N+1)$,\ \ $ S_q(N)\to
-\frac{1}{2}\tilde{\zeta}(N)$, \ \ $T_q(N)\to
-\frac12\tilde{\zeta}(N+1)$  and
 $U_q(N)\,\to
-\frac{\epsilon}{2}\tilde{\zeta}(N)$      
 with $\tilde{\zeta}(N)\equiv\left( N/\tilde{\chi}(N)\right)^{1/2}$,
 so that
 \[   \hspace{-15mm}
 {\cal H}_{q,\epsilon}\buildrel {q\to 1}\over \longrightarrow {\cal H_\epsilon}
= \frac{1}{2(1\!-\!\epsilon^2)}
\left[\left(\tilde{\zeta}(N)c^+\tilde{\zeta}(N+1)c^-
+\tilde{\zeta}(N+1)c^-\tilde{\zeta}(N)c^+ \right) \right.
\]
\[
   \hspace{15mm} \left.  +\,
 \epsilon^2
\left(\tilde{\zeta}(N)c^-\tilde{\zeta}(N+1)c^+ +
 \tilde{\zeta}(N+1)c^+\tilde{\zeta}(N)c^- \right) \right] -
\]
\[ \ \ \
-\, \frac{\epsilon}{2(1\!-\!\epsilon^2)}
\left[\tilde{\zeta}(N)c^+\tilde{\zeta}(N+1)c^+
+\tilde{\zeta}(N+1)c^+\tilde{\zeta}(N)c^+ \right.
\]
\[
\hspace{12mm} \left.
 +\,
\tilde{\zeta}(N)c^-\tilde{\zeta}(N+1)c^- +
 \tilde{\zeta}(N+1)c^-\tilde{\zeta}(N)c^- \right]\, .
\]
 It involves the terms with $c^{\pm} \tilde{\zeta}(N) c^{\pm}$ and
 $c^{\pm} \tilde{\zeta}(N+1)c^{\pm}$ if $\epsilon\neq 0$.
 But if $\epsilon= 0$, in the reduced ($q\!\to\!1$) "quasiparticle"
 Hamiltonian only the first row survives. We thus have
  ${\cal H}_{\epsilon}|_{\epsilon=0}=
  \frac12 \left(\tilde{\zeta}(N)\,c^+\,\tilde{\zeta}(N+1)\,c^-
   + \tilde{\zeta}(N+1)\,c^-\, \tilde{\zeta}(N)\,c^+ \right)$, and use (40),(41).
   As result, the Hamiltonian ${\cal H}_{\epsilon}|_{\epsilon=0}$
   turns into $\frac12 (a^+ a^- + a^- a^+)$.

On the other hand, if in (\ref{H'c+c-}) we set $\epsilon= 0$ and let
$q\neq 1$, the Hamiltonian ${\cal H}_q$ slightly simplifies, but
still contains the terms with
 $c^+...c^+$ and $c^-...c^-$ namely
 \[   
 {\cal H}_{q,\epsilon}|_{\epsilon= 0}
= A_q(N)\zeta(N)c^+\zeta(N)c^+
 + B_q(N)\zeta(N+1)c^-\zeta(N+1)c^-
 \]
 \[  \hspace{14mm}
 +\, C_q(N)\zeta(N)c^+\zeta(N+1)c^-
 + D_q(N)\zeta(N+1)c^-\zeta(N)c^+ .
\]
In this case we have $c^+=\zeta^{-1}(N)a^+$,
 $c^-=\zeta^{-1}(N+1)a^-$, which correspond to the "diagonal"
 DNBT, see the comment after eq.~(41).
  Due to that, the Hamiltonian ${\cal H}_q|_{\epsilon=0}$ goes over
  into that given in (\ref{H_sw}).

Now consider the {\it canonical case} (see Remark 5).
 In this case   
\begin{equation}        \label{S-can}
R^{\rm can.}_q(N)\!=\!\kappa_q^{-1/2}\bigl(A_q(N)-\epsilon\,
D_q(N)\bigr), \ \ \ \
 S^{\rm can.}_q(N)\!=\!\kappa_q^{-1/2}\bigl(\epsilon\, B_q(N)- C_q(N)\bigr),
\end{equation}
\begin{equation}             \label{T-can}
 T^{\rm can.}_q(N)\!=\! \kappa_q^{-1/2}\bigl( \epsilon\, A_q(N)- D_q(N)\bigr) , \ \ \ \
 U^{\rm can.}_q(N)\!=\!\kappa_q^{-1/2} \bigl(B_q(N)-\epsilon\,
 C_q(N)\bigr) ,
\end{equation}
and the Hamiltonian (\ref{H'c+c-}) takes simpler form
$$          
{\cal H}^{\rm can.}= \frac{\kappa^{-1/2}_q}{1-\epsilon^2}
\biggl\{\left[\bigl(R^{\rm can.}_q(N)+\epsilon S^{\rm
can.}_q(N)\bigr)\, c^+ c^+ + \bigl(U^{\rm can.}_q(N)+\epsilon T^{\rm
can.}_q(N)\bigr)\, c^- c^-\right]\biggr. -
$$
\vspace{-4mm}
\[ \biggl.   \hspace{17mm}
 -\left[\bigl(S^{\rm can.}_q(N)+\epsilon R^{\rm can.}_q(N)\bigr)\, c^+c^-
 + \bigl(T^{\rm can.}_q(N)+\epsilon U^{\rm can.}_q(N)\bigr)\,
 c^-c^+\right]\biggr\}
\]            
(recall that GNBTs in this case look as in (45)-(46)\,).

At $q\to 1$, we have $R^{\rm can.}_q(N),\ U^{\rm can.}_q(N)\to
-\frac{\epsilon}{2}$,\ \ $S^{\rm can.}_q(N),\ T^{\rm can.}_q(N)\to
-\frac12$,\ and then come to
$$          
{\cal H}^{\rm can.}|_{q=1}= {\cal H}^{\rm can.} =
\frac{1+\epsilon^2}{2(1-\epsilon^2)} \bigl(c^+ c^- + c^- c^+\bigr) -
\frac{\epsilon}{1-\epsilon^2} \bigl(c^+ c^+ + c^- c^-\bigr) .
$$
With account of (45)-(46) we recover ${\cal H}^{\rm can.} =
H=\frac12 (a^+a^- + a^-a^+)$.

Now rewrite the Hamiltonian in (\ref{H'c+c-}) as
\[
{\cal H}_q =
 \frac{1}{1\!-\!\epsilon^2} \Bigl( R_q(N)\, c^+\,\zeta(N)
 + \epsilon S_q(N)\, c^+\zeta(N+1)\Bigr)\, c^+
\]
\[   \hspace{6mm}
 + \frac{\epsilon}{1\!-\!\epsilon^2} \Bigl(\epsilon T_q(N)\, c^- \, \zeta(N)
 +  U_q(N)\,c^-\zeta(N+1)\Bigr)\,c^-
\]
\[
  \hspace{6mm}   - \frac{\epsilon}{1\!-\!\epsilon^2}
   \Bigl( \epsilon R_q(N)\, c^+\, \zeta(N)
   - S_q(N)\, c^+\, \zeta(N+1)\Bigr) \, c^-
 \]
\begin{equation}                                      \label{H''c+c-}
  \hspace{6mm} - \frac{1}{1\!-\!\epsilon^2} \Bigl( T_q(N) \, c^- \, \zeta(N)
 - \epsilon U_q(N)\, c^-\, \zeta(N+1)\Bigr)\, c^+ .
\end{equation}
 Requiring that the first two lines of the Hamiltonian (with the
 terms $c^+\zeta(...)\,c^+$ and $c^-\zeta(...)\,c^-$) must vanish,
 we {\it impose the following two operator relations}:
\begin{equation}                        \label{R=0}
R_q(N)\, c^+ \zeta(N)\!+
                        \epsilon\, S_q(N)\, c^+ \zeta(N+1) = 0 \ ,
\end{equation}
         \vspace{-7mm}
\begin{equation}                          \label{T=0}
 \epsilon\, T_q(N)\, c^-\zeta(N)\!+\, U_q(N)\, c^- \zeta(N+1) = 0\ .
\end{equation}
 \underline{Remark 6}.
  In the canonical case (see Remark 5), the Hamiltonian (\ref{H''c+c-})
  depends on the multiplier $\kappa_q$.
  However, $\kappa_q$ completely cancels out
  from the constraints (\ref{R=0}) and (\ref{T=0}).
  Second, if we formally put $\epsilon=0$ in the constraints then these lead
  to the relations $R_q(N)=U_q(N)=0$ or respectively $A_q(N)=B_q(N)=0$.
   But the latter can hold only at $q=1$ (i.e. no deformation).
  So for what follows we assume $\epsilon\neq 0$.

  Now let us study the implications of (\ref{R=0}) and (\ref{T=0})
  holding simultaneously. With the use of (\ref{c-}), (\ref{c+}) we express these
  relations in terms of $a^+$ and $a^-$ as
  \[  [x(N)\, R_q(N) + \epsilon_q S_q(N)] a^+
  + \epsilon_q [R_q(N) + \epsilon_q y(N) S_q(N)] a^- =0,
  \]
  \[   [y(N)\, U_q(N) + \epsilon_q T_q(N)] a^-
  + \epsilon_q [\epsilon_q x(N) T_q(N) + U_q(N)] a^+ =0,
  \]
where $x(N)\equiv\zeta^{-1}(N)\zeta(N\!-\!1) $ and
$y(N)\equiv\zeta^{-1}(N\!+\!1)\zeta(N\!+\!2)$.
 Applying the formulas (\ref{a+n}) for the operators $a^+ , a^- $ acting
 in deformed Fock basis we have
 \[
 \sqrt{\Phi_q(n+1)}\, \Bigl(x(n+1)\, R_q(n+1) + \epsilon_q
 S_q(n+1)\Bigr)
\vert n+1\rangle +
\]
\[
 + \epsilon_q \sqrt{\Phi_q(n)}\,
  \Bigl(R_q(n-1) + \epsilon_q y(n-1) S_q(n-1)\Bigr) \vert n-1\rangle =0,
  \]
  \[
  \sqrt{\Phi_q(n+1)}\, \epsilon_q\Bigl( U_q(n+1) + \epsilon_q x(n+1)\,
  T_q(n+1)\Bigr)
\vert n+1\rangle +
\]
\[
 +\sqrt{\Phi_q(n)}\,
  \Bigl(\epsilon_q T_q(n-1) +  y(n-1) U_q(n-1)\Bigr) \vert n-1\rangle  =0 .
  \]
  Vectors $\vert n-1\rangle $ and $\vert n+1\rangle$ are linearly
  independent. Since $\Phi_q(n)\neq 0$, and assuming $\epsilon_q\neq 0$
  ($\epsilon_q=0$ dictates $c^+\sim a ^+$,\ $c^-\sim a ^-$),
   we infer that the equality can be valid only if the following relation
   does hold:
\[
 x(n+1)\, R_q(n+1) + \epsilon_q S_q(n+1) =0, \ \ \ \
 \epsilon_q  y(n-1)\, S_q(n-1) + R_q(n-1)=0 .
\]
These two equations yield: $x(n)= \frac{1}{y(n)}$.
 Taking into account that $y(n)= \frac{1}{x(n+2)}$ we arrive at
 the equality $x(n) = x(n+2)$\ (note that the same can be
 drawn basing on ($\delta'$)\,). The result means the following:
 \[
x(n) = \begin{cases} {\rm e}^{{\rm i} \pi n} = \cos(\pi n) = \left(-1 \right)^n ; \\
   {\rm const}\equiv c_q .
 \end{cases}
 \]
The first option yields
\[
\frac{\zeta(n)}{\zeta(n\!-\!1)}\!=\!
\sqrt{\frac{\Phi_q(n)\,\chi_q(n\!-\!1)}{\Phi_q(n\!-\!1)\,\chi_q(n)}}={\rm
e}^{-{\rm i} \pi n} \ \ \  \Rightarrow\ \ \
 \frac{\Phi_q(n)}{\chi_q(n)}\!=\!\frac{\Phi_q(n\!-\!1)}{\chi_q(n\!-\!1)}
 \ \ \ \Rightarrow\ \ \ \Phi_q(n) \sim \chi_q(n) .
\]
 The second one means that $\zeta(n)\equiv\zeta_q(n)$ admits the
form $\zeta_q(n)=c_q^{-n}\zeta(0)$ where $\zeta(0)$ can be set as
$\zeta(0)=1$, in view of footnote on page 10.
 Thus, $\chi_q(n)=c_q q^n \Phi_q(n)$, which at $q\!\neq\!1$,\
$q^{n}\!\neq\!1$ looks as a "modification" of the canonical case
(\ref{kappa}).

 The latter analysis shows  we must examine more
 thoroughly the case and consequences of the canonical DNBT.
 That will be done in the next Section.

\vspace{-2mm}
   \section{Diagonalized Hamiltonian for free quasi-particles}

 Requiring that (\ref{R=0}) and (\ref{T=0}) do hold, the Hamiltonian
 takes the form
\[   \hspace{-15mm}{\cal \tilde{H}}_q =
  \frac{1}{\epsilon^2\!-\!1} \Bigl(\epsilon \, R_q(N)c^+
\zeta(N)\, c^- + S_q(N)c^+ \zeta(N+1)\, c^-\Bigr)
  \]
       \vspace{-6mm}
 \begin{equation}    \hspace{-8mm}
 + \frac{1}{\epsilon^2\!-\!1} \Bigl(T_q(N) c^-
\zeta(N)\, c^+ + \epsilon \, U_q(N) c^- \zeta(N+1)\, c^+\Bigr) \
 \end{equation}
 and can be also given through only two operator functions
 say $S_q(N)$ and $T_q(N)$:
 \begin{equation}                            \label{c+c-}
     \hspace{-22mm}
  {\cal \tilde{H}}_q = - \Bigl(S_q(N)\, c^+\, \zeta(N+1)\, c^- +
 T_q(N)\, c^-\, \zeta(N)\, c^+ \Bigr).
 \end{equation}
 This is quasi-free (i.e. depending on the products $c^+...\,c^-$ and
 $c^-...\,c^+$) Hamiltonian for quasi-particles which are most general
 ($\chi$-)deformed bosons whose operators obey (\ref{chi}).

 It is hardly possible to diagonalize ${\cal \tilde{H}}_q$ in (\ref{c+c-})
 with general deformation function $\chi(N)$, and below we consider the case
 of canonical DNBT (see Remark 5).
Note that in this canonical case
$\zeta(N+1)=\zeta(N)=\kappa_q^{-1/2}$, and then we have
 \begin{equation}    \hspace{-22mm}
  {\cal \tilde{H}}_q \rightarrow {\cal \tilde{H}}^{\rm can.}_q
  = - \kappa_q^{-1/2} \Bigl(S^{c}_q(N)\, c^+\, c^- +
 T^{c}_q(N)\, c^-\, c^+ \Bigr).
 \end{equation}
In the limit $q\to 1$, we have $\kappa_q\to 1$,\ $S^{c}_q(N)\to -
\frac12, \ T^{c}_q(N)\to - \frac12$, and the familiar Hamiltonian
${\cal H}=\frac12(a^+a^- + a^-a^+ )$ is recovered.

With account of Eq.~(37) we come to the following Hermitian
Hamiltonian for free quasi-particles expressed as a function of the
excitation number operator:
\begin{equation}                     \label{H-chi}
         \hspace{-22mm}
 {\cal \tilde{H}}^{\rm can.}_q
  = \frac{- 1}{\sqrt{\kappa_q}}\, \Bigl(S^{c}_q(N)\,\chi(N) +
 T^{c}_q(N)\,\chi(N+1) \Bigr).
\end{equation}
The spectrum of this Hamiltonian will be found in the next
Subsection.

 \subsection{Eigenvalues of the quasi-particle Hamiltonian}

 In this section we examine the distinguished case of {\it
"canonical"} GNBT, see Remark 5, for which most advanced results can
be achieved. Recall that in the case of usual (linear) Bogoliubov
transformations, the term canonical refers to those transformations
which preserve commutation relations.
  In the present deformed situation we impose slightly weaker requirement,
  that the DSF resulting after the {\it "canonical"} GNBT being applied
  is equal to the initial DSF $\Phi(N)$ upto the multiplier
  $\kappa_q^{-1}$, depending on deformation parameter(s)
  and not depending on $N$, and such that $\lim_{q\to1}\kappa_q =1\,$ is satisfied.
 In this (canonical) case eqs.~(\ref{c-})-(\ref{c+}) simplify and
 reduce to eqs.~(\ref{can})-(\ref{can-inverse})\ 
(recall the action formulas (7)-(8) for $a^+$ and $a^-$).

From (\ref{H-chi}) we infer the result for the spectrum,
\[
E_n \equiv  \langle n\vert {\cal \tilde{H}}^{\rm can.}_q \vert
n\rangle
= - \sqrt{\kappa_q}\, \Bigl(S^{c}_q(n)\,\Phi_q(n) +
 T^{c}_q(n)\,\Phi_q(n+1) \Bigr)
\]
where the relation (\ref{kappa}) has been taken into account.
    Recalling the explicit form of $S^{c}_q(n)$ and
$T^{c}_q(n)$ from (\ref{S-can}) and (\ref{T-can}) we finally obtain:
\[    \hspace{-10mm}
 E_n =\frac{\, \ q^{3 n}}{4} \Bigl[
 \Bigl(q^{1/2}+q^{-1/2}\Bigr)\left(q^{-3/2}\Phi_q(n)+q^{3/2}\Phi_q(n+1)\right)\Bigr.
 \]
 \begin{equation}                   \label{spec}
 \hspace{1mm}
 - \Bigl.\epsilon_q \Bigl(q^{1/2}-q^{-1/2}\Bigr)\Bigl(q^{3}\Phi_q(n)+q^{-3}
 \Phi_q(n+1)\Bigr) \Bigr]
\end{equation}
 where the expression for $\Phi_q(N)$ is given in (\ref{Phi(n)}).

This is our main result. It can also be equivalently presented as 
\begin{equation}                   \label{main}
E_n =E_n(q)= \frac{\, \ q^{3 n}}{4}
 \Bigl(V_q\, \Phi_q(n)+ W_q\,\Phi_q(n+1) \Bigr)
\end{equation}
 where
 \vspace{-1mm}
 \[
  V_q = q^{- \frac32}\Bigl(q^{1/2}+q^{-1/2}\Bigr) -
 \epsilon_q\, q^3 \Bigl(q^{1/2}-q^{-1/2}\Bigr) ,
 \]
 \vspace{-2mm}
 \[
 W_q = q^\frac32\Bigl(q^{1/2}+q^{-1/2}\Bigr) -
 \epsilon_q\, q^{-3} \Bigl(q^{1/2}-q^{-1/2}\Bigr) .
\]
Note that the ground state energy essentially depends on $q$: \
$E_0=E_0(q)=\frac{W_q}{2 q (1+q^2)}$.

It only remains to examine the properties of $\epsilon_q$ and
admissible $q$.

\subsection{Function $\epsilon_q$ and admissible values of
deformation parameter}

So let us explore the explicit     
form and main properties of $\epsilon=\epsilon(q)\equiv \epsilon_q$
in the expression for energy eigenvalues $E_n=E_q(n)$. To this end
we apply the relations (49)-(50).  In the canonical case these take
the form
\begin{equation}         
\left(R^{c}_q(N) + \epsilon_q\, S^{c}_q(N)\right)\, c^+  = 0 \ ,
\ \ \ \ \              
 \left(\epsilon_q\, T^{c}_q(N)\, +\, U^{c}_q(N)\right)\, c^- = 0\
\end{equation}
where $R^{c}_q(N),\ S^{c}_q(N),\ T^{c}_q(N),\ U^{c}_q(N)$ are the
same as in eqs. (48)-(49). For validity of these relations for all
$N$, the two-term sums in each bracket should equal to zero.
 Having acted on Fock basis states $\vert n\rangle$ this gives
 \[
\epsilon^2_q\,A_q - \epsilon_q\,\left(C_q + D_q \right)+B_q=0 ,\ \ \
\ \ \epsilon^2_q\, B_q - \epsilon_q\,\left(C_q + D_q \right)+A_q=0
 \]
where $A_q,\ B_q,\ C_q,\ D_q$ are given in (27), and it is meant
that $q\neq 0 \ {\rm nor}\ 1$ (the latter implies $\epsilon_q\neq 0$
). Requiring compatibility of solutions of the two relations, we
find the following possibilities:
\begin{equation}                   \label{AB}
     \hspace{-26mm}
A_q B_q \neq 0 \ \ \ C_q + D_q \neq 0\, : \ \ \ \epsilon_q=\frac{A_q
+ B_q}{C_q + D_q}\, ;
 \end{equation}
 \begin{equation}                   \label{A+B}
A_q + B_q \neq 0 \, : \ \ \ {\tilde\epsilon}^2_q (A_q + B_q) - 2
{\tilde\epsilon}_q (C_q + D_q) + A_q + B_q = 0 .
  \end{equation}

\underline{Case ({\bf A})}.\  This corresponds to eq.~(\ref{AB}).
 Since both $A_q B_q \neq 0$, \ $C_q + D_q \neq 0$\,
 and $q\neq 0 \ {\rm nor}\ 1$, for positive $q$ from
 (\ref{AB}) we deduce:
 \begin{equation}                   \label{eps1}
 \epsilon_q =q^{-\frac12}\, \frac{(q-1)(q^2+1)(q^2+q^{-2}-1)}{(q+1)^2
 (q+q^{-1}-1)}= q^{-\frac32}\, \frac{(q-1)(q^6+1)}{(q+1)(q^3+1)} \equiv r^{-1}_q.
   \end{equation}
  Admissible values of $q$ are such that $|\epsilon_q|<1$ and $\epsilon_q\neq
   0$. Evaluation gives the result: \ $-1<\epsilon_q<0$ for $q$ in
   the interval $1 >q > 0.4739142 (\approx 0.4739)$, and \ $0<\epsilon_q<1$ for $q$ in
   the range $1 <q < 2.11008657 (\approx 2.11)$.

   If $q\to q^{-1}$ we have $\epsilon_q\to \epsilon_{q^{-1}}=-\epsilon_q$. That gives the
equality $W_q=V_{q^{-1}}$ for the coefficients in formula
(\ref{main}) for $E_n$. Now it can be rewritten as
$E(n)=E_q(n)=\frac{q^{3n}}{4} \left[ V_q\, \Phi_q(n)+ V_{q^{-1}}\,
\Phi_q(n+1)\right]\equiv E^{(1)}_q(n) + E^{(2)}_q(n)$.
 The obtained ranges of admitted values of $q$ for the intervals
 $-1<\epsilon_q<0$ and $0<\epsilon_q<1$ are interchangeable for the
 respective intervals of $\epsilon_{q^{-1}}$.
  This means that admissible $q$ cannot be common for $V_q$ and
  $V^{-1}_q$. Therefore, with possible physical application(s) in mind,
  we have to retain in the initial Hamiltonian (see eq.~(27))
  only one term -- either that leading to $E^{(1)}_q(n)$ or leading to
  $E^{(2)}_q(n)$.

  \underline{Case ({\bf B})},\ linked with equation (\ref{A+B}).
   Using the quantity $r_q$ from (\ref{eps1}), we infer
   \begin{equation}                   \label{eps2}
 {\tilde\epsilon}_q =  r_q \pm \sqrt{r_q^2 -1} \, , \ \ \ q>0 ,
     \ \ \ q\neq 1\, .  
   \end{equation}
Clearly, to the condition ${\tilde\epsilon}_q\neq\pm 1$ there
corresponds the requirement $r_q\neq\pm 1$. Real solution of
(\ref{eps2}) is possible:\ 1) at $r_q> 1$;\ 2) at $r_q < -1$.
Moreover, if $r_q> 1$ then $-1< {\tilde\epsilon}_q\equiv
{\tilde\epsilon}^-_q=r_q - \sqrt{r_q^2 -1}<1$, and if $r_q < -1$
then $-1< {\tilde\epsilon}_q\equiv {\tilde\epsilon}^+_q=r_q +
\sqrt{r_q^2 -1}<1$. Hence it remains to clarify for which values of
$q>0$ the conditions 1) and 2) are valid. The analysis yields that
1) $r_q> 1$ if $1 <q < 2.11008657 (\approx 2.11)$, and 2) $r_q < -1$
when $1 >q > 0.4739142 (\approx 0.4739)$.

If $q\to q^{-1}$ we have $r_q\to r_{q^{-1}}=-r_q$ and thus
${\tilde\epsilon}^{\pm}_q\to -{\tilde\epsilon}^{\mp}_q$ so that,
unlike in previous Case {\bf A}, now we have
${\tilde\epsilon}^{\pm}_{q^{-1}}=-\Bigl(r_q\mp \sqrt{r^2_q-1}\Bigr)=
-{\tilde\epsilon}^{\mp}_q\neq -{\tilde\epsilon}^{\pm}_q$.
 Therefore in this case $W_q\neq V_{q^{-1}}$, and in the expression for
 $E_q(n)$ we may operate, unlike Case {\bf A},r with the both two terms,
 the one with $\Phi_q(n)$ and the othe with $\Phi_q(n+1)$.

Thus the obtained two expressions ${\tilde\epsilon}^{\pm}_q\neq 0$
belonging to the interval
 $-1 < {\tilde\epsilon}^{\pm}_q<1$ and satisfying eq.~(65) provide
 most general solution for the problem of spectrum of the deformed
Hamiltonian eq.~(27) (transformed into eq.~(54) and then into
eq.~(56)). Herein, the range of admissible values of deformation
parameter $q$ covers the interval $0.4739 < q < 2.11$
(with $q=1$ dropped).

In Figs. 1 and 2 we plot the energy spectrum function (\ref{main})
at different values of deformation parameter.
 We observe the nontrivial (namely non-monotonic) behavior of $E_q(n)$
 as function of $n$. Such type of behavior suggests~\cite{Plethora}
 a possibility of pairwise {\it accidental} degeneracy of chosen
 (pairs of) energy levels at certain $q$.
  \vspace{-2mm}
\begin{figure}[h]
\includegraphics[width = 0.55\linewidth]{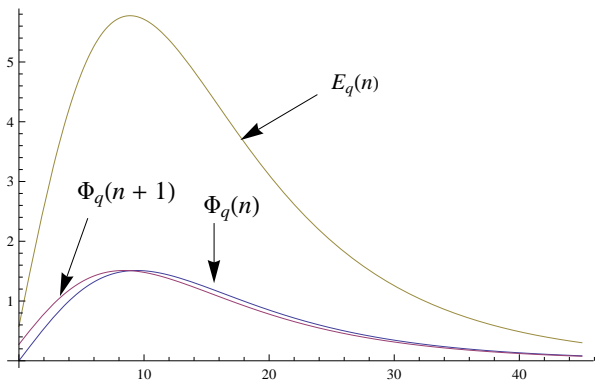}  %
\caption{The functions $\Phi_q(n)$,\ $\Phi_q(n+1)$, and $E_q(n)$
versus excitation number $n$, at fixed $q=1.1$.}
  \label{fig1}
\end{figure}
  \vspace{-3mm}
\begin{figure}[h]
\includegraphics[width = 0.55\linewidth]{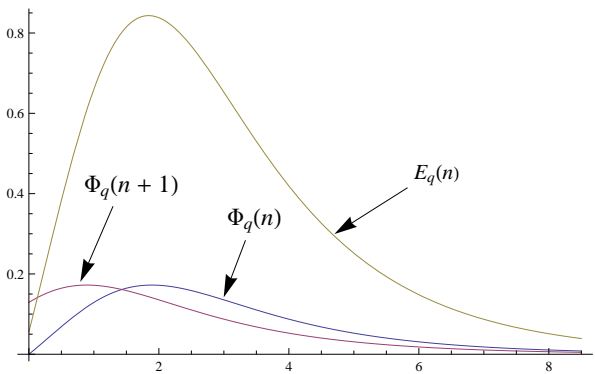}  %
\caption{The functions $\Phi_q(n)$,\ $\Phi_q(n+1)$, and $E_q(n)$
versus excitation number $n$, at fixed $q=0.59$.}
  \label{fig1}
\end{figure}

 \vspace{-2mm}
 \section*{Concluding remarks}
In this paper we have constructed the Hermitian Hamiltonian from
non-Hermitian ingredients -- $\eta_X(N)$-pseudo-Hermitian position
operator $X$ and $\eta_P(N)$-pseudo-Hermitian momentum operator $P$,
and explored its properties.
  Because of high (in fact non-polynomial) nonlinearity of the Hamiltonian,
  we have developed the generalized nonlinear Bogoliubov transformation
  which essentially differ from the usual ones (which are linear and involve
  $c$-number coefficients).
  In the distinguished case of canonical GNBT, by definition, 
  the statistics determined by the structure function remains unchanged
  up to a multiplier $\kappa_q$, that is,
   $\Phi(N)$ $\to$ $\chi(N)=\kappa_q \Phi(N)$. 
  A natural choice is to set $\kappa_q=1$.

  When the GNBT has been applied with the goal to diagonalize the Hamiltonian,
  we have inferred the constraints that are basically different from
  the case of usual Bogoliubov transformations:\ indeed
  the constraints (\ref{R=0}) and (\ref{T=0}), based on GNBT and aimed as
  the tools for diagonalization, {\it are the operator ones}.
      It would be of interest to try to extract some physical
    consequences of these relations.

  Our second main result is the energy spectrum (\ref{spec}) of the
  Hamiltonian diagonalized explicitly in the case of canonical GNBT.
   In this connection, we have analyzed in detail the ranges of
   admissible values of the deformation parameter $q$.
   The plots given in Fig.~1 and 2 show, for few chosen values of $q$,
   the nontrivial behavior of the spectrum as a function
   of the quantum number $n$.

  \vspace{-1mm}
\section*{Acknowledgement}

This work was partially supported by the Special Program, Project
No. 0117U000240, of Department of Physics and Astronomy of National
Academy of Sciences of Ukraine.

\end{document}